# Retardance of lab grown diamond substrates as a function of thickness: momentum-drift random walk model

*Thanh Tran [1,*], Phuong Vo [2], Thomas Sheppard [1], Timothy Grotjohn [1,3], Paul Quayle [1]*

[1] Great Lakes Crystal Technologies, Inc., 4942 Dawn Ave, East Lansing, Michigan 48823, United States

[2] Department of Statistics, Michigan State University, East Lansing, Michigan 48824, United States

[3] Department of Electrical and Computer Engineering, Michigan State University, East Lansing, Michigan 48824, United States

[*] Author for correspondence, email: thanh@glcrystal.com

**Abstract**

This work studies the correlation between mean retardance and thickness of diamond substrates grown homoepitaxially via microwave plasma-enhanced chemical vapor deposition (MPCVD). We measure the retardance of a diamond substrate in two orientations: perpendicular and parallel to the growth direction. Our experimental results demonstrate that the correlation between mean retardance and thickness differs for these orientations. When measured perpendicular to the growth direction, the mean retardance is approximately proportional to the square root of the substrate thickness. In contrast, when measured parallel to the growth direction, we observe a generally higher mean retardance and an approximately linear correlation with thickness. This anisotropy arises not from differences in stress magnitude but from differences in the interlayer correlation of the principal stress axes, as evidenced by correlation coefficients between the azimuth angles of consecutive layers in the diamond crystal. To simulate the integrated retardance of diamond wafers, we propose a two-dimensional random walk model with momentum drift, which captures the diamond crystal tendency to preserve the azimuth angle across the samples. By optimizing the momentum factor, we show that the model can closely match experimental data. The momentum factor is found higher along the growth direction, which is consistent with the calculated correlation coefficients. Furthermore, both the model and experiments indicate that retardance-to-thickness ratios of thin samples converge toward similar base retardances in both orientations. These findings establish a quantitative framework for interpreting birefringence in diamond substrates, with implications for material selection and development in thermal management, quantum sensing, high-power electronics, and optical applications.

## Introduction

Diamond has remarkable materials properties and it promises to enable next-generation applications in thermal management and advanced packaging of AI and power electronic chips [1, 2], high-power electronics [3, 4, 5, 6, 7] , and quantum sensing [8, 9]. In addition to common challenges such as producing and processing large-area diamond substrates and achieving n-type doping, manufacturing high-quality diamond substrates is essential for advancing these applications [10, 11, 12, 13, 14]. Quantitative birefringence is a non-destructive technique, and it provides valuable insights into the stress caused by defects within a diamond sample. In the context of nitrogen-vacancy (NV) centers in diamond, this lattice stress negatively impacts the $T_2^*$ spin coherence time of the NV centers. Stress introduces inhomogeneous broadening of the NV center spin dynamics. This leads to a reduction in the spin coherence time of NV centers and limits their use in quantum sensing and information applications [15, 16, 17, 18]. Therefore, quantitative birefringence analysis serves as a reliable method for identifying high-quality diamonds suitable for quantum sensing and other applications where low stress is essential. Moreover, efficient and effective characterization of diamond quality is of paramount importance to quality control, and continued research and development.

Various methods are available for evaluating the quality of diamond substrates, including X-ray topography (XRT), X-ray diffraction (XRD) for full width at half maximum analysis, etched pit counting, Raman spectroscopy, electron channeling contrast imaging (ECCI), and birefringence measurement [19, 20, 21, 22, 23, 24]. Among these, birefringence offers a quick and cost-effective solution that provides full-field data within the material [25, 26, 27].

As light enters a birefringent material, it will split along slow and fast axes. Birefringence, denoted as $\Delta n$, is given by the difference between the refractive indices along the slow axis and the fast axis [28, 29]. Formally, it is expressed in equation (1)

$$\Delta n = n_1 - n_2 \tag{1}$$

where $n_1$ and $n_2$ are the refractive indices measured along the fast and slow axes, respectively.

In the birefringence measurements, we do not directly measure birefringence but rather its byproducts: phase retardation, which represents the phase difference between the fast and slow axes of a birefringent material for light passing through it, and the azimuth angle, which is the angle that the fast axis (or slow axis) of a birefringent material makes with a reference axis. The method for measuring these quantities is thoroughly explained by Yokoyama et al. [30].

In photoelasticity of optically isotropic materials, the birefringence $\Delta n$ is modeled as proportional to the difference between two principal stresses, $(\sigma_1 - \sigma_2)$ via the stress-optic coefficient $C$. Specifically, $\Delta n$ is shown in equation (2)

$$\Delta n = C(\sigma_1 - \sigma_2) \tag{2}$$

where $\sigma_1$ and $\sigma_2$ denote the principal stresses, and $C$ is a material-specific parameter [31, 32, 33, 34].

According to Pinto et al. and Srinivasan et al., [29, 35]:

$$C = n^3(q_{11} - q_{12})/2 \quad (3)$$

where $n = 2.42$ is the refractive index of diamond at 633 nm in this case, and $q_{ij}$ are piezo-optic constants of diamond. The value of $q_{11} - q_{12}$ is calculated to be $7.8 \times 10^{-13}\ Pa^{-1}$ according to Ramachandran [33]. Thus,

$$C_{diamond} \approx 5.5 \times 10^{-12}\ Pa^{-1}. \quad (4)$$

*Figure 1* shows that the histogram of retardance of a diamond sample is well approximated by a Rayleigh distribution. As shown in equation 2, the retardance is proportional to the difference between principal stresses, $\sigma_1 - \sigma_2$. Following to Yokoyama et al. [30], the stress magnitude defined as in equation (5)

$$\sigma_1 - \sigma_2 = \left( \left(\sigma_{xx} - \sigma_{yy}\right)^2 + 4\sigma_{xy}^2 \right)^{\frac{1}{2}} \quad (5)$$

which follows a Rayleigh distribution, when $\left(\sigma_{xx} - \sigma_{yy}\right)$ and $2\sigma_{xy}$ are independent, normally distributed with mean zero, and share the same standard deviation [36]. If these assumptions do not hold, the distribution becomes a generalized Rayleigh distribution, known as Hoyt distribution, where the shape parameter is determined by the ratio of the standard deviations of $2\sigma_{xy}$ and $(\sigma_{xx} - \sigma_{yy})$ [37]. Due to the evolution of integrated retardance along thickness, we expect to observe deviations in the measured retardance distribution from the idea Rayleigh distribution.

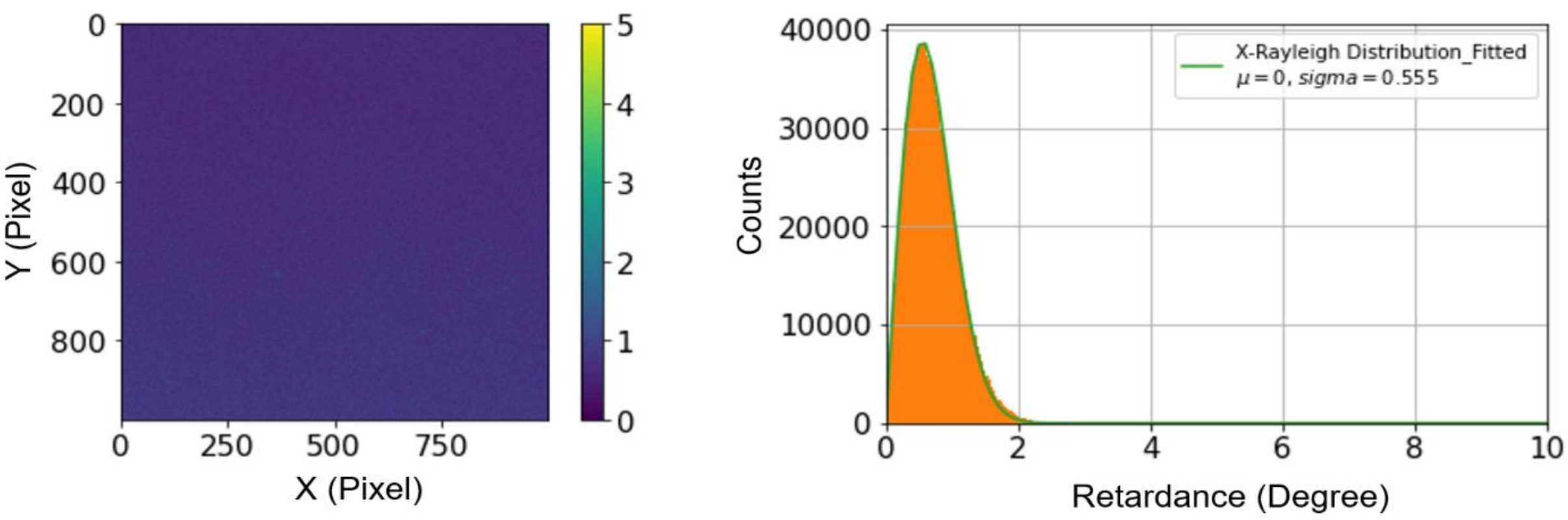

*Figure 1: Example of retardance map of a high-quality type IIa HPHT diamond substrate and its corresponding histogram diagram fitted with a Rayleigh distribution.*

A common normalization assumes a linear relationship between the measured retardance, Γ, and the birefringence, $\Delta n$, such that $\Delta n = \frac{\Gamma}{d}$, where $d$ denotes the thickness of the material at the measurement location [21, 22]. However, this assumption holds true only when the azimuth angle remains consistent across layers, meaning the fast and slow axes direction are the same across the sample. An example is the case of birefringent materials, which have the fast and slow axes determined by crystal orientation [38]. It is worth noting that in photoelasticity of optically isotropic materials like diamond, principal stresses axes directly coincide with the fast and slow axes. Therefore, they are interchangeable in this study of diamond without affecting the result of the study.

The linear assumption does not hold true in the case of diamond. Because of cubic lattice structure and $C_4$ rotational symmetry, diamonds are optically isotropic [39, 27]. Therefore, diamond without internal stress has no birefringence. Birefringence in diamond is strain/stress-induced, arising from defects such as dislocations, inclusions, and fractures [40, 41, 42, 43]. These randomness of defect's location cause variations in fast axes from one layer to another. Therefore, the linear assumption is not appliable in this case. To use birefringence measurement for evaluating quality of diamond substrates, the correlation between measured retardance and sample thickness must be well-understood, enabling normalization of measurements across samples with varying thicknesses. This motivates this study, which aims to investigate the correlation between the thickness and retardance of diamond samples. To study the correlation, we also propose and investigate a momentum-drift random walk model capturing two variables representing base retardance and the tendency to maintain principal stress axes.

**Experimental study**

To perform the experiment, a diamond layer was grown on a commercial type II CVD substrate using the MPCVD method. The top surface of the diamond sample was aligned with the (100) lattice plane, with an off angle of about 3 degrees. The reactor used for growing diamonds was the bell-jar type developed at Michigan State University [44, 45]. To minimize global stress, during the growth, a constant amount of nitrogen was introduced, and the temperature was maintained at a fixed value. A temperature variation of less than 5 °C across the sample was observed using a thermal camera. After the growth, the thickness of the sample increased from 1 mm to approximately 2.3 mm. Two areas were selected and cut from the substrate for different investigations (*Figure 2* (a)).

Sample 1, which is called Cross, is cut and polished to have the starting dimensions of 2x2.3x10 in mm. The sample then was flipped 90 degrees (*Figure 2* (c)) to study the retardance in the direction perpendicular to the growth direction. Sample 2, called Wafer, was laser-planarized and polished to remove the as-grown surface and had the starting dimensions of 3.5x3.5x2 mm. This sample was used to investigate retardance along the growth direction (*Figure 2* (b)). Both samples were thinned by sequentially removing material layer by layer using a laser cutter. After each laser cut, we gently polished the sample using mechanical polishing technique to remove subsurface damage and residual stress caused by the cutting process. The influence of stress caused by subsurface damage from processing steps such as laser cutting and polishing is negligible. This conclusion is supported by the observation that, after applying the same processing steps to a high-quality, high-pressure high-temperature (HPHT) diamond sample, the measured retardance is nearly at the background noise level, approximately 0.7 degrees, measured without a sample. In addition, in this study, the contribution of background noise was statistically removed from the mean retardance under the assumption that there is no correlation in azimuth angle between background noise and samples.

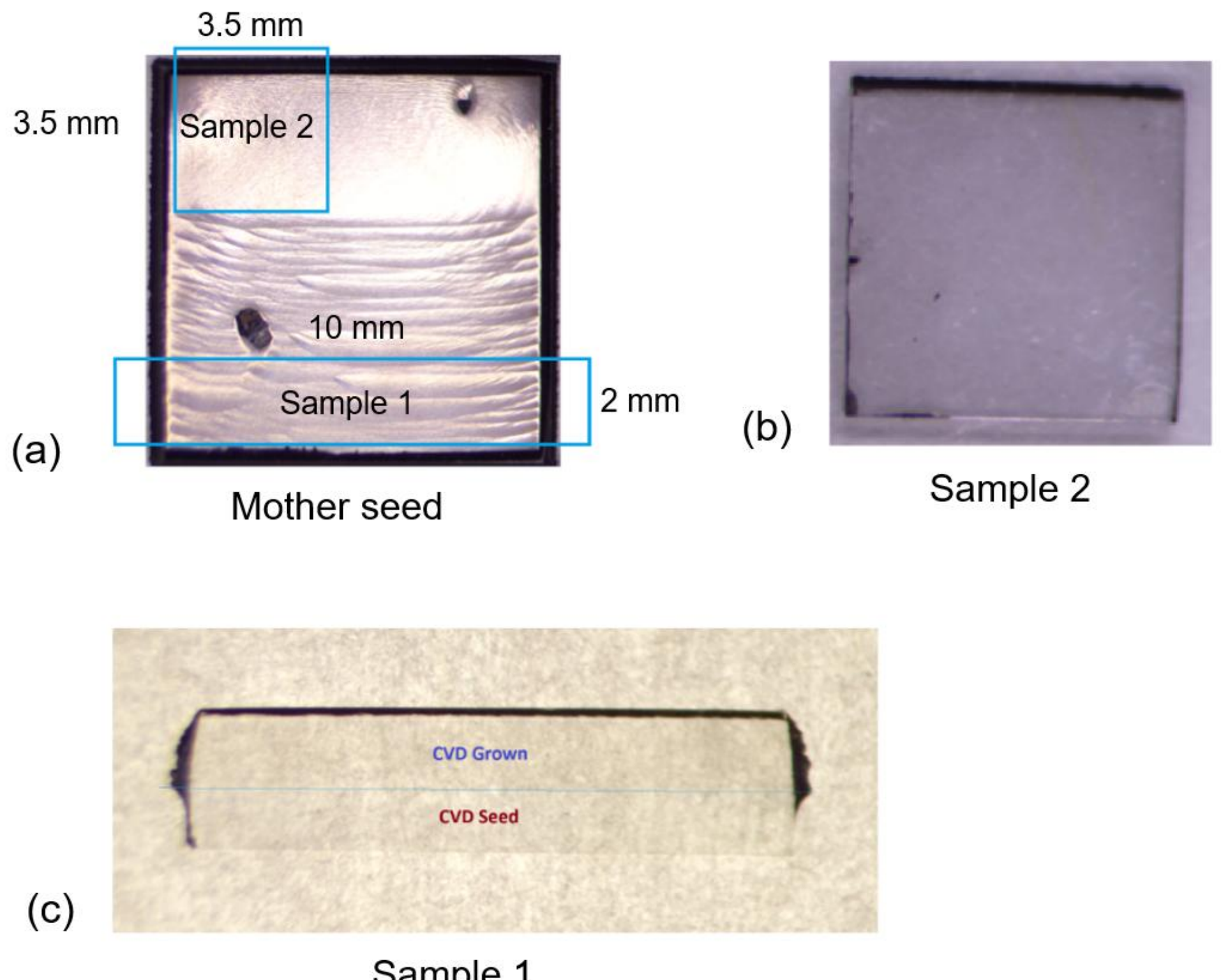

*Figure 2: (a) Mother seed and its cutting layout showing sample 1(Cross) and sample 2 (Wafer) used for retardance study, (b) Sample 2 after being cutting out of the mother seed, and (c) Sample 1 after being cut out of the mother seed.*

*Figure 3* illustrates the processing procedures for Wafer and Cross samples. The retardance maps were recorded for the series of Cross and Wafer samples at varying thickness. During the thinning process, some Thin Wafers and Slices cut from the Wafer and Cross samples, respectively, were successfully polished and collected (*Figure 3*). These additional pieces were analyzed to study the correlation between retardance and azimuth angle, the direction of the principal stresses [46], within diamond's layers in the two different orientations. In this study, quantitative birefringence was measured using CM501 Birefringence Imaging Microscope with a laser wavelength of approximately 632.8 nm. This system measures linear retardance, defined as the phase difference between two linear polarization components in the sample, and the orientation of the fast axis, reported here as the azimuth angle. The interchange of retardance units from degree to nm is through equation (6):

$$\Gamma_{nm} = \Gamma_{degree} \times \frac{632.8}{360} \tag{6}$$

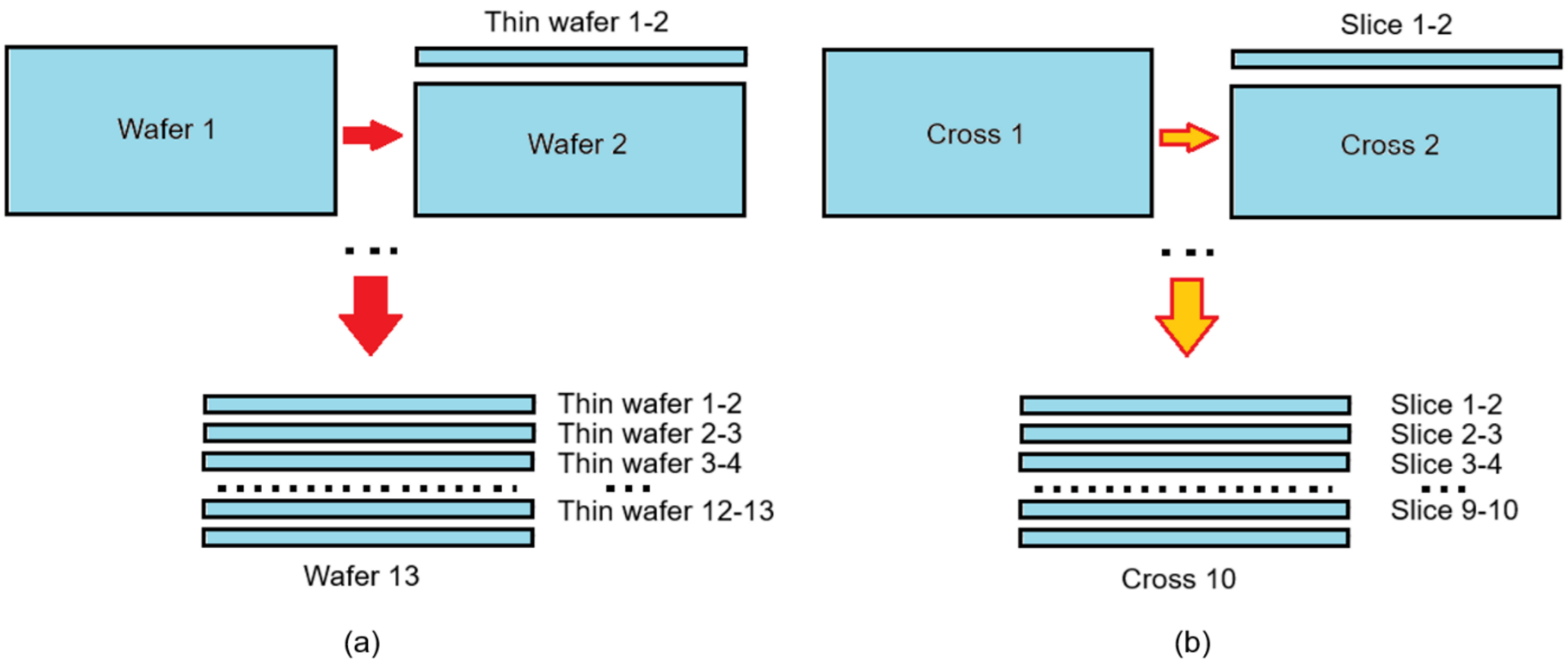

*Figure 3: Diagram showing processing procedure for (a) Sample 2 (Wafer) and (b) Sample 1 (Cross). Data is collected in each step before moving to next step.*

**Results and Discussion**

*Figure 4* (a) and (b) shows retardance maps of the samples Cross 1 and Wafer 1, which are the starting samples of the Cross and Wafer samples respectively. Despite both having thicknesses of approximately 2 mm and being cut from the same parent seed, Wafer 1 exhibits significantly higher birefringence than Cross 1. This anisotropy has been reported in several studies. In 2009, Friel et al. observed significantly lower birefringence, estimated by linearly scaling retardance with thickness, when measured in the orientation perpendicular to the growth direction ($\Delta n < 1 \times 10^{-6}$) compared with the parallel orientation ($\Delta n < 4 \times 10^{-5}$) [22]. The report attributed the reduction in birefringence in the perpendicular direction to the "hopping" of light in and out of strain fields generated by dislocations. While this explanation may apply to specific regions where dislocation bundles are present, it does not fully describe the behavior across the entire sample. Moreover, it fails to account for the non-linear relationship between retardance and thickness observed later in this study.

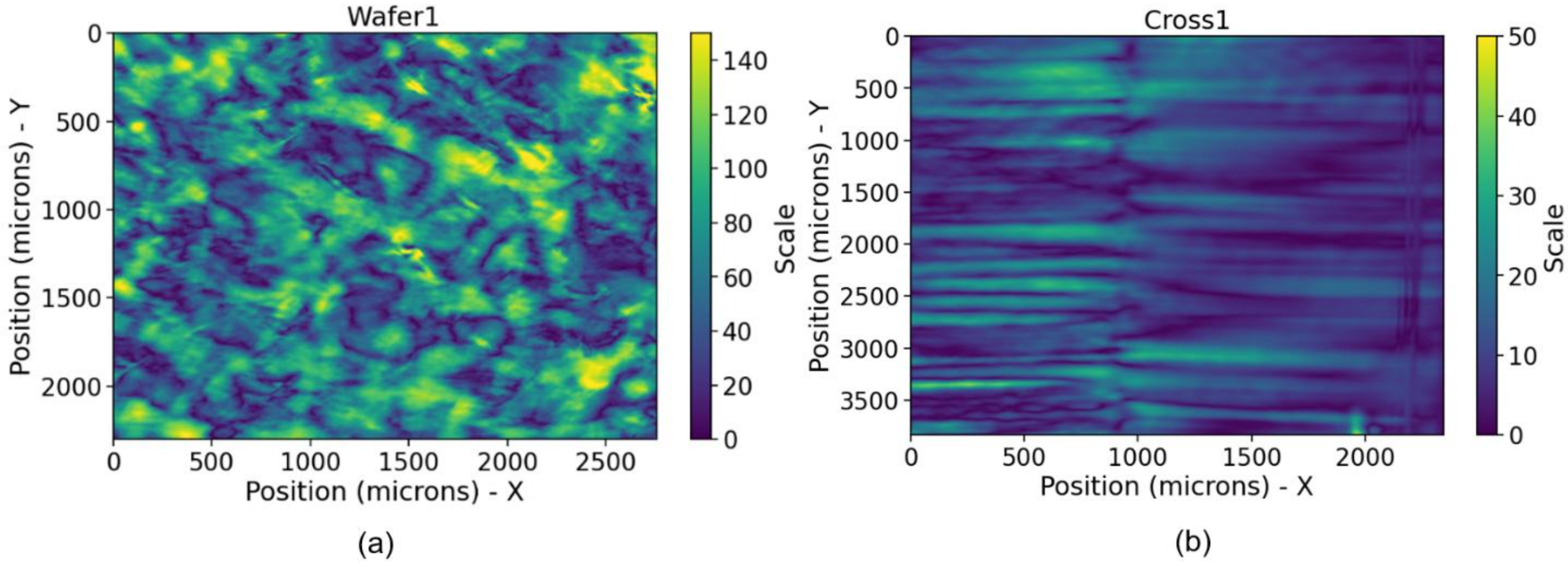

*Figure 4: Retardance map of (a) Wafer 1 (left) and (b) Cross 1 (right)*

In the case of the sample Wafer 1, the retardance high spots appear to arrange in clusters, likely due to dislocation bundles propagating along the [100] direction [47]. This observation aligns with *Figure 4* (b), where bright strips running along the [100] direction are attributable by strain fields of dislocation bundles. Additionally, the interface between the seed layer (left side) and the grown layer (right side) is clearly visible. According to Tsubouchi et al., the initial growth layer is a significant source of mixed dislocations [41]. This leads to the disruption of stress patterns in the perpendicular direction of viewing and makes it visible in the retardance maps.

*Figure 5* (a) summarizes the mean retardance as a function of thickness for Cross samples, revealing a non-linear correlation. *Table S1* (Supporting information) provides birefringence maps and histogram for selected samples with varying thicknesses, offering additional insights into the variation in retardance. This observation suggests that the principal stress axes do not persist throughout layers of samples. *Figure S3* (Supporting information) displays retardance maps of aligned Slice samples, further demonstrating the low correlation between layers.

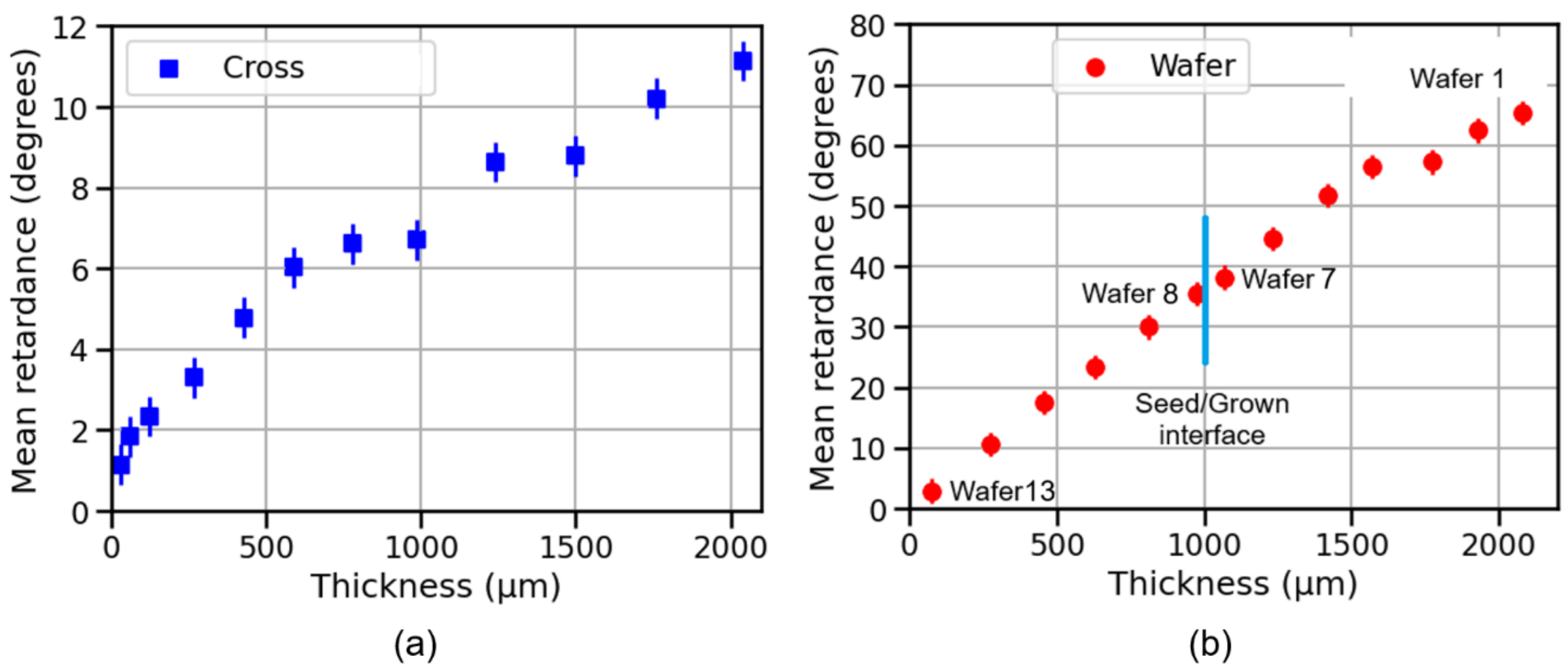

*Figure 5: (a) Thickness dependance of mean retardance of Sample 1 (Cross) and (b) Thickness dependances of mean retardance of Sample 2 (Wafer).*

*Figure 5* (b) illustrates the mean retardance as a function of thickness for Wafer samples. The graph shows a near-linear correlation, suggesting a high correlation of azimuth angle between layers. *Figure S1* (Supporting information) shows the retardance maps and *Figure S2* (Supporting information) shows the azimuth maps of various aligned thin Wafers. These maps demonstrate a strong correlation between the upper and lower layers, particularly when the layers are in close proximity.

*Figure 6* (a) and (b) shows Pearson's product moment correlation coefficient matrix of azimuth angle between aligned Thin Wafers and Slices samples. This coefficient, which ranges from -1 to 1, illustrates that the strength and direction of a linear relationship between two variables gives the statistical relationship between two random variables [48, 49]. The persistence of the principal stress axes between consecutive layers can be quantified using the Pearson correlation which is calculated in equation (7).

$$r = \frac{\Sigma_{i=1}^{N}\{(X_i - \bar{X})(Y_i - \bar{Y})\}}{\sqrt{\Sigma_{i=1}^{N}\,(X_i - \bar{X})^2}\,\sqrt{\Sigma_{i=1}^{N}\,(Y_i - \bar{Y})^2}} \tag{7}$$

where $\bar{X}$ and $\bar{Y}$ are the sample means of the two variables $X$ and $Y$, repsectively. Here $X_i$ and $Y_i$ denotes the azimuth angles of the two layers. Larger values of $r$ indicate stronger correlation between the two layers, implying greater persistence of the tendency for the principal stress axes.

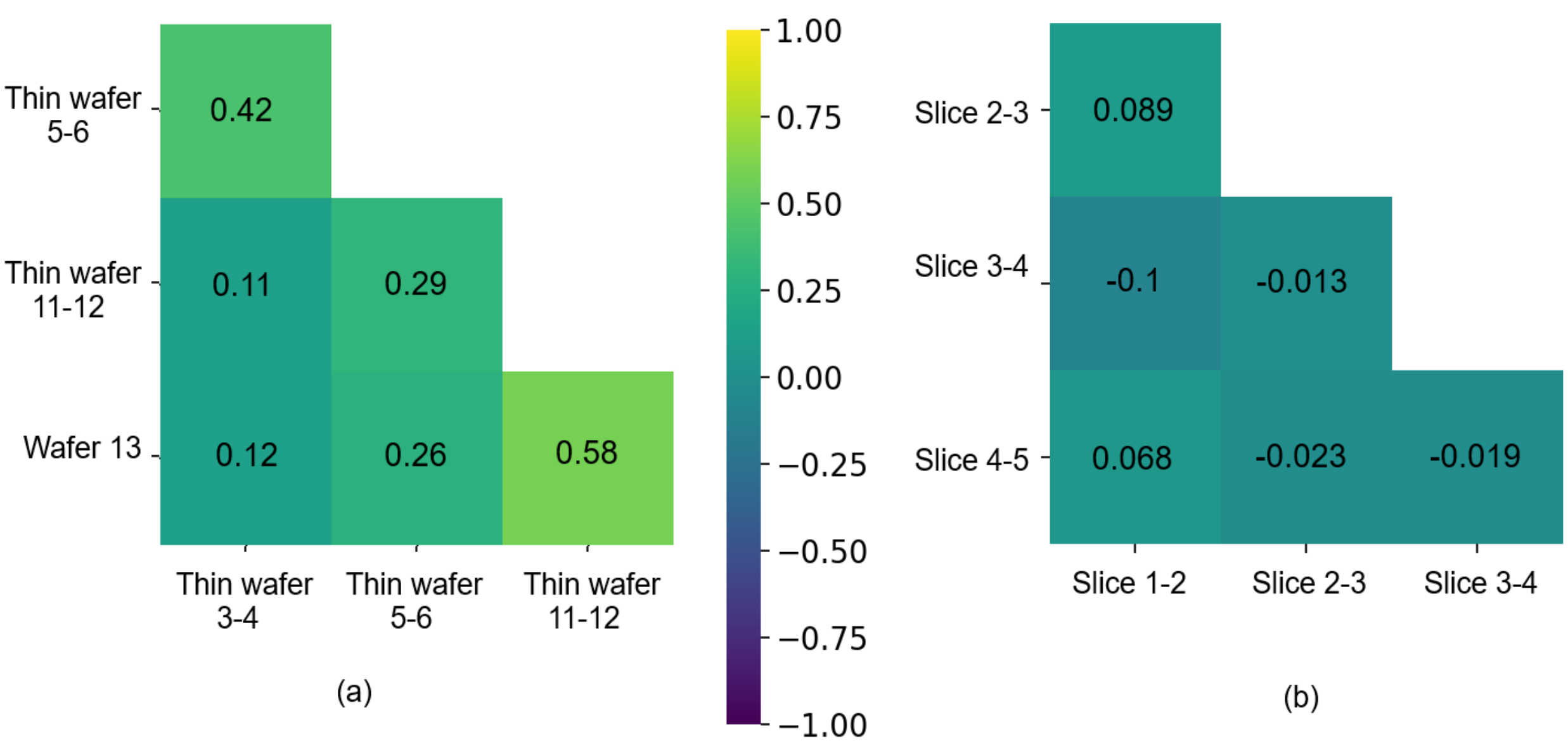

*Figure 6: Azimuth angle correlation between some available (a) Thin Wafers and (b) Slices*

*Figure 6* (a) demonstrates a significant correlation of azimuth angles among the Thin Wafers. This strong correlation is likely due to dislocations that are highly aligned along the growth direction. In CVD growth, such dislocations are typically inclined randomly from the [001] direction within a range of 7°-12° [50]. These anisotropic defects produce strain fields that are independent of their propagation orientation [29]. As a result, the orientation of the principal stress axes is likely preserved throughout the thickness of the sample. Thin Wafers within grown layers or seed layers tend to have higher correlation than Thin Wafers from different layers. The closer the layers, the higher the correlation. There is a noticeable correlation between thin wafer 11-12 and thin wafer 5-6. This shows the propagation of threading dislocation from the seed to the grown layer that is observed in several works reported [51, 52, 53, 54]. There are some newly formed dislocations observed in thin wafer 3-4 and 5-6. Most of the newly formed defects are from the thin wafer 5-6, in the very beginning of the grown layer. These defects add randomness and lower correlation between seed and grown layers. Therefore, the correlation is positive but less than the correlation between layers from the same seed layer or grown layer. In the case of Slices as shown in *Figure 6* (b), the correlation of azimuth angle between samples is negligible, demonstrating the low tendency to maintain principal stress axes.

With the principal stress axes vary from layer to layer differently in the two orientations, we expect to see different behaviors of retardance. To simplify the situation, we consider two linear retarders with retardance of $\delta_1$ and $\delta_2$ stacking together. The Jones matrix for the first layer having fast axis aligned at angle 0 degrees from the coordinate system is

$$J_1 = \begin{bmatrix} e^{i\delta_1/2} & 0 \\ 0 & e^{-i\delta_1/2} \end{bmatrix} \quad (8)$$

The Jones matrix for the second layer, $J_2$, with retardance $\delta_2$ and fast axis rotated by $x^o$ relative to the fast axis of the first layer, can be expressed as

$$J_2 = R(-x). \begin{bmatrix} e^{i\delta_2/2} & 0 \\ 0 & e^{-i\delta_2/2} \end{bmatrix} . R(x) \quad (9)$$

where $R(x) = \begin{bmatrix} \cos(x) & -\sin(x) \\ \sin(x) & \cos(x) \end{bmatrix}$ is the rotation matrix that transforms the coordinate system by an angle $x$ [55].

Consequently, the compound Jones matrix of both layers is:

$$J_{total} = J_1 . J_2 \quad (10)$$

The eigenvalues of the compound Jones matrix, $J_{total}$, determine the overall retardance $\delta_{total}$. By performing the matrix multiplication and solving for the eigenvalues, the exact combined retardance is given by equation (11) [56]:

$$\cos\left(\frac{\delta_{total}}{2}\right) = \cos\left(\frac{\delta_1}{2}\right)\cos\left(\frac{\delta_2}{2}\right) - \sin\left(\frac{\delta_1}{2}\right)\sin\left(\frac{\delta_2}{2}\right)\cos(2x) \quad (11)$$

When $\delta_2 \ll \delta_1$, the formula simplifies to:

$$\delta_{total} \approx \delta_1 + \delta_2 \cos(2x) \quad (12)$$

where $\delta_2 \cos(2x)$ represents the rotated modulation on a 2D plane of the second layer.

Let the fast-axis rotated alignment of layer $k$ be $x_k \sim Unif[0, \pi)$. Then the combined angle of $2x_k \sim Unif[0, 2\pi)$, which is equivalent to a conventional 2D random walk model. Consequently, it is natural to study retardance–thickness behavior using a 2D random walk model.

In the context of the random walk model, the correlation between successive azimuth angles can be interpreted as a tendency to maintain the orientation from the previous step. To study this effect, we suggest a momentum-drift random walk model, expressed as follows:

$$\Gamma = \left|\sum_{j=1}^{N} S_j\right| = \Gamma_{base}\left|\sum_{j=1}^{N} e^{i\theta_j}\right| = \Gamma_{base}\left|\sum_{j=1}^{N} \frac{Me^{i\theta_{j-1}} + R_j}{\left\|Me^{i\theta_{j-1}} + R_j\right\|}\right| \quad (13)$$

where:

- $N$ denotes the number of steps
- $S_j$ denotes the step j of the walk
- $M$ is the momentum component factor, quantifying the tendency to maintain azimuth angle
- $\Gamma_{base}$ is base retardance of sample, which will be explained further in the next paragraph
- $R_j$ is the random component of the step $j$, sampled uniformly from the set $\{1, -1, i, -i\}$ corresponding to right, left, up, and down movements, respectively
- $\theta_j$ is the orientation in the complex plan of step $j$.

Unlike a conventional random walk model, where each step is completely independent and random, the momentum-drift random walk model introduced here includes a component that retains a portion of the previous momentum. This addition represents the tendency of the azimuth angle to be maintained. In the simulation of this work, the step denotes the increment of retardance after adding a layer of 1 $\mu m$ thick of diamond. Base retardance is the mean retardance of the 1 $\mu m$ thick diamond layer. In the ideal case, it has a Rayleigh distribution. *Figure S7*, Supporting Information, illustrates an example of the probability density function of an ideal base retardance map. Besides the momentum component, the model has a random component that is the same as a conventional random walk model, having magnitude of one and four random directions. The ratio between the momentum and random components controls the degree of directional persistence. In this model, the random component has magnitude of one. Therefore, the ratio is equal to momentum component factors. The magnitude of the step is the base retardance. It is worth noting that momentum component factors will change when the thickness for determining base retardance changes. Therefore, the momentum specified in this study is calculated specifically for the given thickness step of 1 $\mu m$. The graphical illustration of the momentum-drift random walk model is presented in *Figure 7* (a). When the momentum component is zero, the model reduces to a conventional random walk. Conversely, when the momentum component dominates (approaching infinity), the movement becomes effectively deterministic, with the expected retardance growing linearly with the number of steps.

To study the behavior of this model, we implemented a computational simulation and calculated the expected displacement as a function of the step number ($N$) with a step size $\Gamma_{base} = 1$ averaging over 2000 independent random walks. *Figure 7* (b) illustrates the resulting behavior of the expected displacement for different momentum component factors: 0, 5, 20, and 100. As expected, the trend of the expected value transitions from being proportional to the square root of the number of steps ($\propto \sqrt{n}$) to being linearly proportional to the number of steps ($\propto n$) as the momentum component increases.

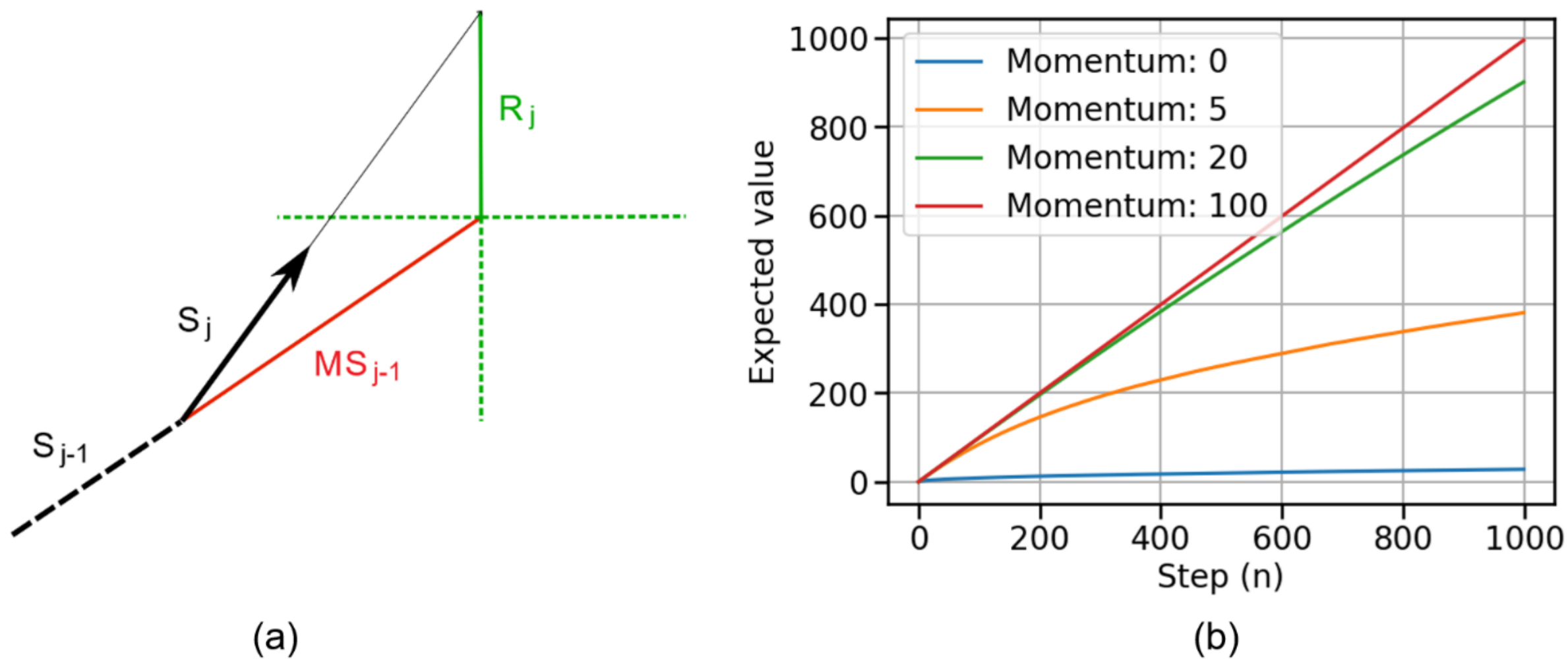

*Figure 7: (a) Schematic diagram of momentum-drift random walk model. The green component represents the random component, which can be either* $i$, $-i$, $1$, *or* $-1$, *and the red component represent the momentum component. (b) A survey study of expected values as functions of number of steps with unit step size (base retardance equal one) at four different momentum component factors: 0, 5, 20, and 100.*

To simulate actual retardance, we need to determine base retardance. The correlation study and the preliminary simulation results suggest that Cross samples exhibit lower momentum components compared to Wafer samples. Additionally, we observed that when the sample is thin, the correlation between retardance and thickness tends to be closer to linear. To estimate the base retardance, Wafer 13 (*Figure S1*), which is 74 $\mu m$ thick and has a mean retardance of 2.89 degrees, is used. This yields an approximate base retardance of 0.039 degrees $\mu m^{-1}$. In the case of Cross samples, the thinnest sample we produced is the Slice 8-9 having thickness of 30 $\mu m$ and a mean retardance of 1.15 degrees (*Table 1*). This leads to a base retardance of 0.038 degrees $\mu m^{-1}$, which is close to that of Wafer 13. This indicates that the Wafer and Slice samples base retardances likely to converge to the same value. Accounting for the momentum factor, the actual base retardance is expected to be higher than the estimated values. In this work, base retardance of 0.04 degrees $\mu m^{-1}$ is chosen for both orientations to find momentum factors.

Two fittings were performed, and we found momentum factors of 2.4 and 20 for Cross and Wafer samples, respectively (*Figure 8* (a)). These are the result of scanning momentum factors to optimize the R-squared value with a given base retardance of 0.04 (*Figure S8* and *Figure S9*, Supporting Information). The fitted lines show good agreement with the experimental data with R-squared values for the two fittings of 0.990 and 0.995, respectively. In Wafer samples, the grown CVD layer shows high turbulence at the grown layer due to newly created defects during the growth, as shown in Thin Wafer 5-6 and Thin Wafer 3-4 (*Figure S1*, Supporting Information). This explains the deviation of experimental and fitting, especially in the grown layer. Within the seed layer only, the R-squared value is up to 0.9999 (*Figure S10*, Supporting Information).

*Table 1: Thickness and mean retardance of Thin Wafers cut out of Cross samples used for analyzing the correlation and of Slice 8-9 sample for getting base retardance.*

| Sample | Thickness (µm) | Mean retardance (Degrees) | Base retardance (Degrees/µm) |
|---|---|---|---|
| Thin wafer 3-4 | 80 | 3.52 | 0.044 |
| Thin wafer 5-6 | 74 | 3.30 | 0.045 |
| Thin wafer 11-12 | 65 | 2.73 | 0.042 |
| Wafer 13 | 74 | 2.92 | 0.039 |
| Slice 8-9 | 30 | 1.15 | 0.038 |

The above fitting uses the same momentum and base for both seed and grown layers. A more accurate fitting will take into consideration different bases and momentum factors for each layer. An example is shown in *Figure 8* (b). The new fitting has a better R-squared value of 0.997 by assigning base retardance for the seed and grown layers of 0.04 and 0.045 with momentum factors of 20 and 15, respectively. The base retardance of 0.045 for the grown layer was chosen by referencing base retardance of the Thin Wafer 3-4 and 5-6 in *Table 1*. After obtaining the base retardances of the two layers and the momentum factor of the seed layer of 20, a momentum factor of 15 for the grown layer was determined by optimizing R-squared value as plotted in *Figure S11*, Supporting Information. This fitting indicates that the grown layer of this substrate has higher stress and lower momentum compared to the seed layer. This agrees well with experimental observations. *Figure 6* shows that the correlation between the azimuth angle of Thin Wafers 3-4 and 5-6, which is 0.46, is lower than that between Thin Wafer 11-12 and Wafer 13, which is 0.58. *Table 1* illustrates properties of Thin Wafers and Wafer 13, and it points out that Thin Wafers 3-4 (0.044) and 5-6 (0.045) have higher base retardance than that of Thin Wafer 11-12 (0.042) and Wafer 13 (0.039). This is reflected by new stress areas observed in retardance maps of Thin Wafers 4-5 and 5-6 (*Figure S1*, Supporting Information), which adds more stress into the layers and increases the weight of random component, subsequently reducing the momentum component. A more rigorous fitting could be carried out by optimizing the R-squared value for both base retardances and momentum factors. While deriving an explicit analytical solution for the momentum-drift random walk model would facilitate this fitting process, it is beyond the scope of this work.

The current analysis and high R-square fittings reveal that both Cross and Wafer samples share a similar base retardance, while Wafer samples exhibit a higher momentum component resulting in a higher retardance for Wafer samples. This opens the possibility to use base retardance as an indicator, which is highly independent of crystal orientation, to evaluate the quality of CVD diamonds. One way to find the base retardance is to make samples thin enough to observe a linear dependance between thickness and mean retardance. At that thickness, retardance approaches a form close to $t \times R_{base}$, where $t$ is thickness and $R_{base}$ is the base retardance. In general, the lower the momentum factor, the thinner the sample is needed to get reliable base retardance as shown in *Figure 7* (b).

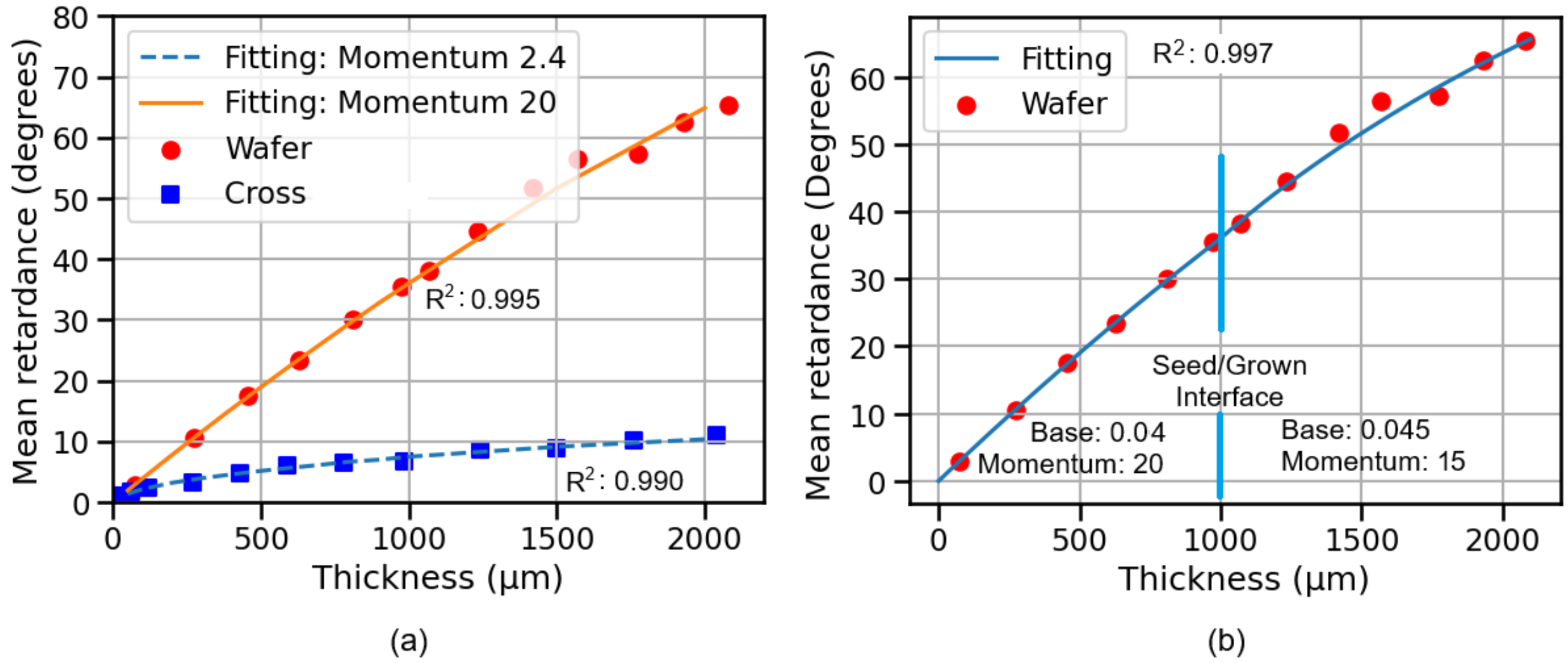

*Figure 8: (a) Experimental retardance mean of Wafer and Cross samples together with their momentum-drift random walk fittings at momentum factor of 2.4 and 20, respectively. (b) A better Wafer samples fitting considering differences in base retardance mean and momentum of seed and grown layers.*

An intuitive interpretation of the momentum factor is that it quantifies how many layers, perceived as a diffusion length, are required for the materials to lose memory of its initial fast-axis orientation. A momentum factor of 0 in the model means a retardance diffusion length of $1\ \mu m$. At this condition,

$$\Gamma_t = \Gamma_{base} \times \sqrt{\frac{\pi}{4} \times t} = 0.04 \times \sqrt{\frac{\pi}{4} \times t}\,. \tag{14}$$

For the orientation perpendicular to the growth direction, a momentum factor of 2.4 corresponds to a retardance diffusion length of more than $1 um$. The experimental retardance in this case can be well fitted by

$$\Gamma = 0.237 \times \sqrt{t} \approx 0.04 \times \sqrt{\frac{\pi}{4} \times 45 \times t}, \tag{15}$$

where $t$ is the thickness of diamond wafers in $\mu m$. Thus, 45 $\mu m$ can be regarded as the effective diffusion length of retardance in this orientation.

This diffusion length is strongly related to density and distribution of dislocations and dislocation bundles parallel to the growth orientation, as these are the dominant sources of long-range strain in diamond crystals. The “hopping” in and out of dislocation strain fields, as described by Friel et al. (2009), can be modeled as steps in a two-dimensional random walk [22]. Considering 45 $\mu m$ as the average distance between dislocations, the dislocation density can be estimated as

$$\frac{1}{45 \times 45\ \mu m^2} \approx 5 \times 10^4\ cm^{-2}. \tag{16}$$

For the optical path parallel to the growth direction, a momentum factor of 20 corresponds to a retardance diffusion length of around 3000 $\mu m$. The retardance exhibits a much longer diffusion length, since the light propagates along dislocations that are aligned with the growth direction. The standard random-walk model, using the diffusion length as the step size, can account for retardance of samples much thicker than the diffusion length. However, in this case, for samples with thickness comparable to or smaller than the diffusion length, a momentum-drift random-walk model is required. This framework naturally captures the transition from the linear regime to the diffusive regime, providing the consistent description across different thickness scales.

In the linear regime, the birefringence of a sample can be obtained from the fundamental relation:

$$\Delta n = \frac{\Gamma_n}{d}. \tag{17}$$

For the sample in this study, with a base retardance of 0.04 degrees, the birefringence is calculated as

$$\Delta n = \frac{\Gamma_n}{d} = \frac{0.04\ (degree) \times \frac{632.8\ (nm)}{360\ (degree)}}{1 \times 10^3 (nm)} = 7.0 \times 10^{-5}. \tag{18}$$

Accordingly, using the stress-optic coefficient $C = 5.5 \times 10^{-12}\ Pa^{-1}$ from the equation (4), the principal stress difference is

$$\sigma_1 - \sigma_2 = \frac{\Delta n}{C} = \frac{7.0\times10^{-5}}{5.5\times10^{-12}\ Pa^{-1}} \approx 12.7\ MPa. \tag{19}$$

Thus, the principal stress difference $\sigma_1 - \sigma_2$ is expected to follow a Rayleigh distribution with a mean of approximately 12.7 MPa. Assuming that dislocations are the main sources of the birefringence, a rough estimate of dislocation density can be made based on the study by Pinto et al., where 80 dislocations within an $800 \times 800\ \mu m^2$ area lead to an average stress of about 3 MPa [29]. By scaling linearly to the stress of 12.7 MPa in this study, the estimated dislocation density is

$$\frac{80}{800 \times 800\ um^2} \times \frac{12.7\ MPa}{3\ MPa} \approx 5 \times 10^4\ cm^{-2}. \tag{20}$$

This estimated value and the value in equation 15 agree with each other and they fall within the range observed in etch pit studies conducted on similar samples, although those results are not reported here.

Beyond diamond, base retardance can be adapted to become a useful matrix to evaluate quality of materials other than diamond such as SiC, GaN, AlN, and Silicon [57, 58, 59, 60, 61, 62]. In the case of birefringent materials, it is possible that some further steps are needed to remove linear retardance and extract photo-elastic retardance caused by internal stress from defects within the crystals.

## Conclusion

This study investigates the relationship between retardance and thickness in lab-grown diamond substrates along two principal orientations: parallel and perpendicular to the growth direction. Experimental results show that retardance measured perpendicular to the growth direction is generally lower than that measured along the growth direction. However, this difference does not indicate lower internal stress in the perpendicular orientation. Instead, it reflects differing degrees of correlation in principal stress axes between layers. Specifically, retardance in the perpendicular direction scales closer with the square root of thickness, consistent with a low inter-layer correlation—akin to a random walk. In contrast, retardance in the parallel direction scales more linearly, suggesting a higher degree of alignment in the principal stress axes across layers. This is evidenced by azimuth angle correlation between consecutive layers in both orientations. Furthermore, this behavior is captured and explained by a proposed physics-informed momentum-drift random walk model, which incorporates both random and momentum-preserving components. Simulations of this model align well with experimental data, indicating that Wafer samples (parallel orientation) exhibit higher momentum components than Cross samples (perpendicular orientation). Importantly, the study shows that the thinner the sample, the more similar the retardance values between the orientations. This convergence at low thickness suggests that, for diamond, base retardance allows for more meaningful comparisons of diamond quality across samples, regardless of crystal orientation.

These findings provide a valuable framework for interpreting quantitative birefringence measurements in lab-grown diamonds and enhance our ability to evaluate internal stress and crystalline quality across substrates of varying thickness and orientation. This is particularly relevant for advanced applications in quantum sensing, high-power electronics, and optical devices, where internal stress can significantly impact performance. It is worth noting that the knowledge in this study can be extended and applied to non-diamond materials such as SiC, GaN, AlN, and Silicon.

## Acknowledgement

This work was supported by the Air Force Research Laboratory award no. FA239424CB029

# Supporting information

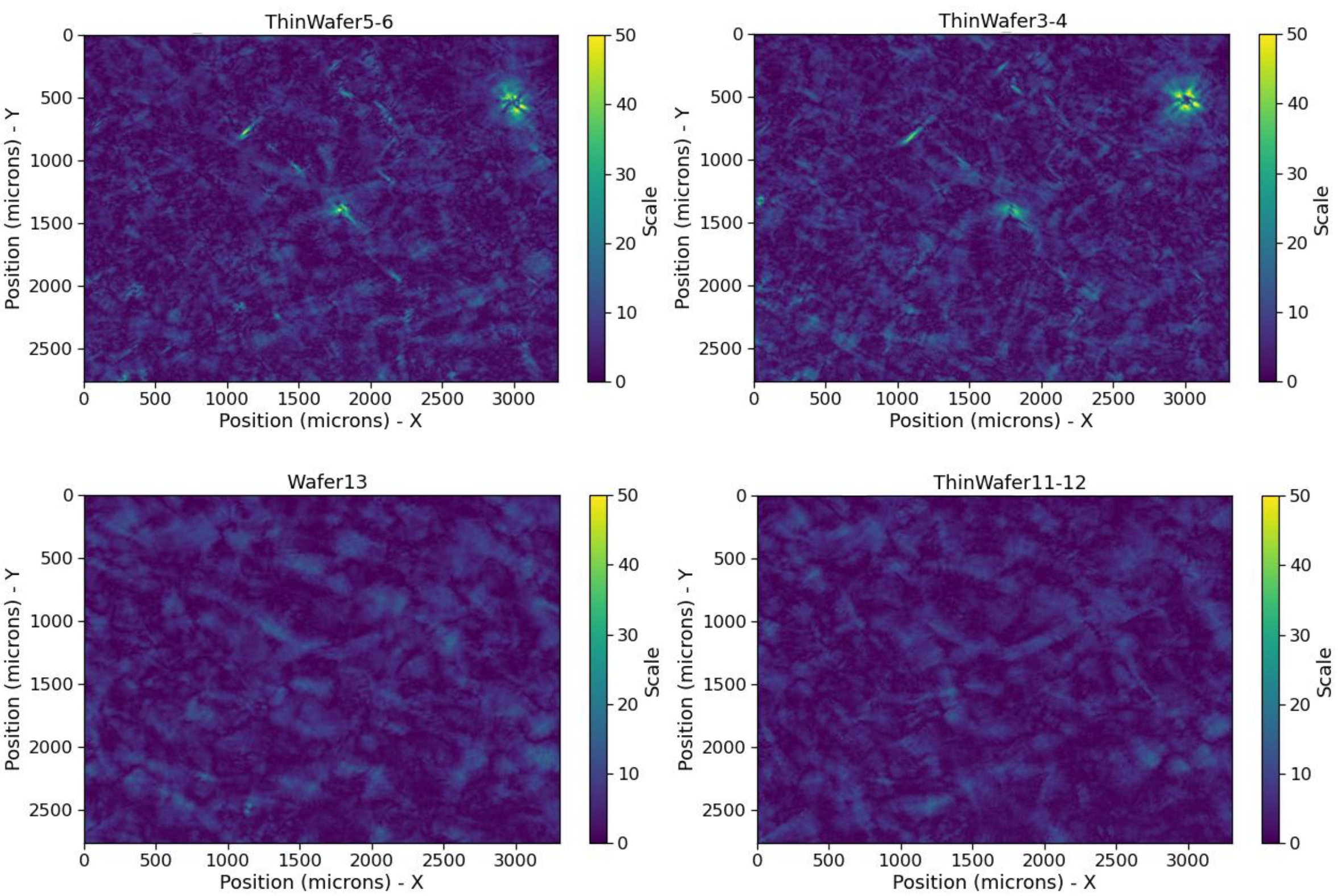

*Figure S1: Retardance maps of the Thin Wafers used for calculating correlation coefficients between layers perpendicular to the growth direction*

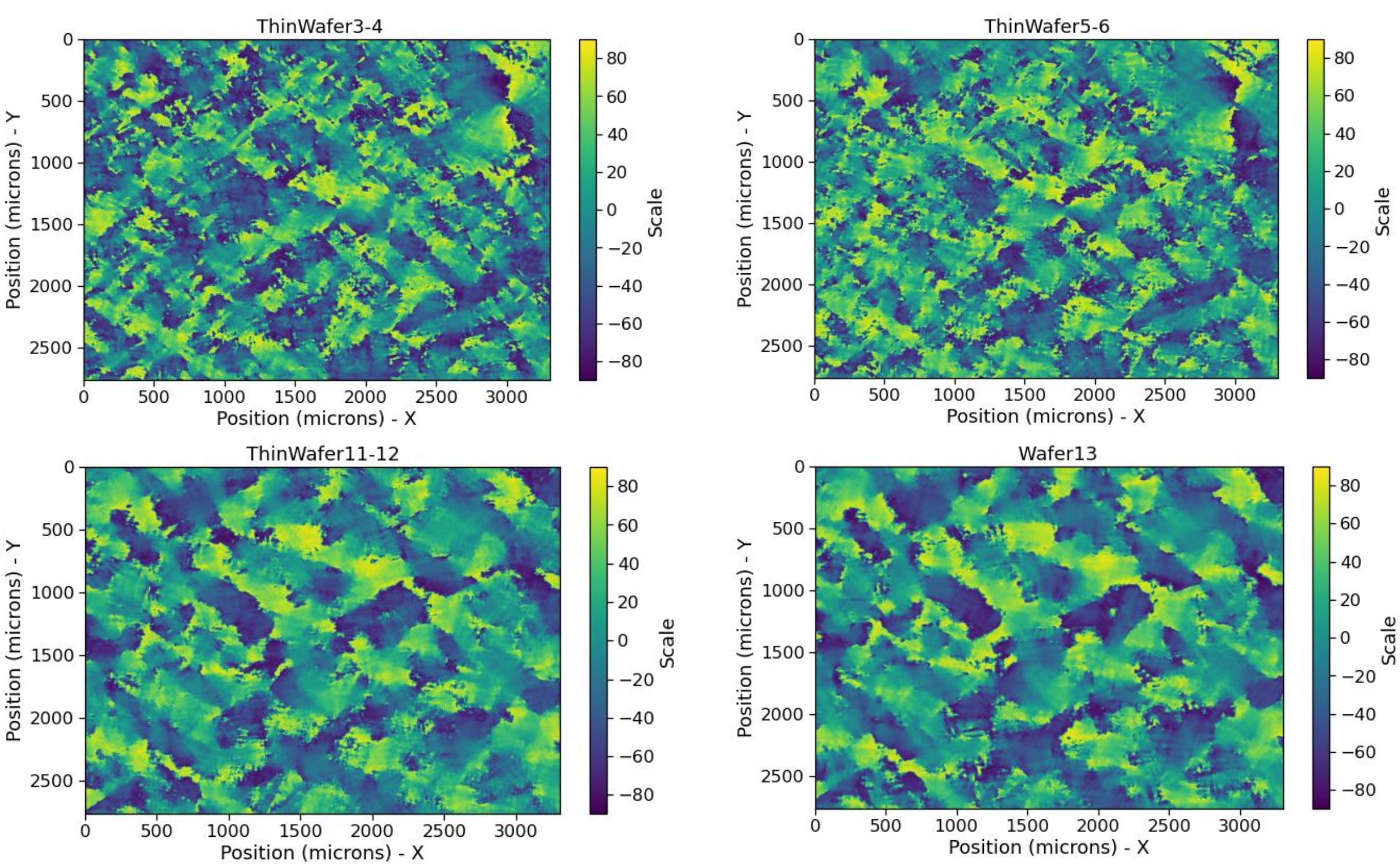

*Figure S2: Azimuth maps of the Thin Wafers used for calculating correlation coefficients between layers perpendicular to the growth direction*

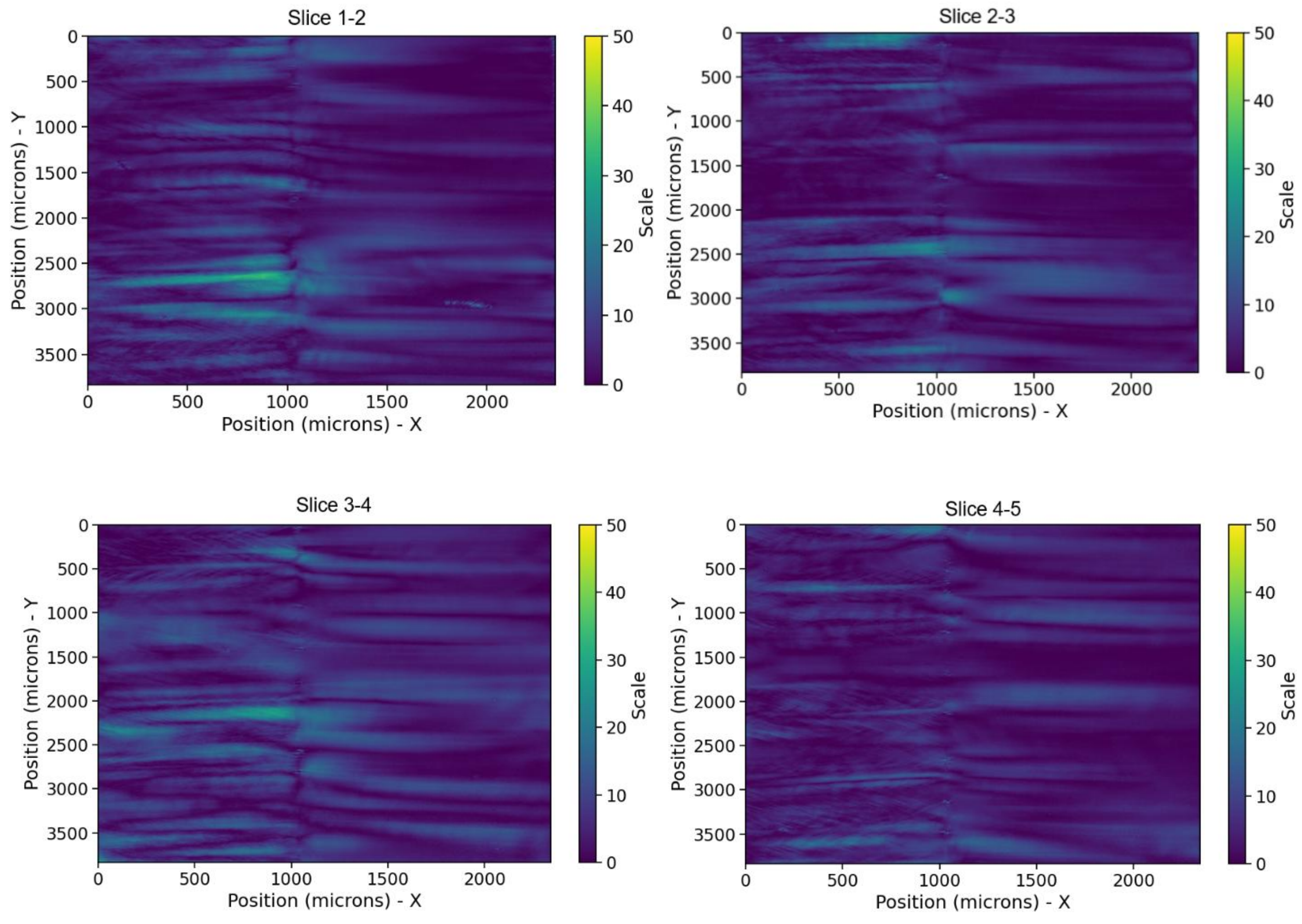

*Figure S3: Retardance maps of the Slices used for calculating correlation coefficients between layers parallel to the growth direction*

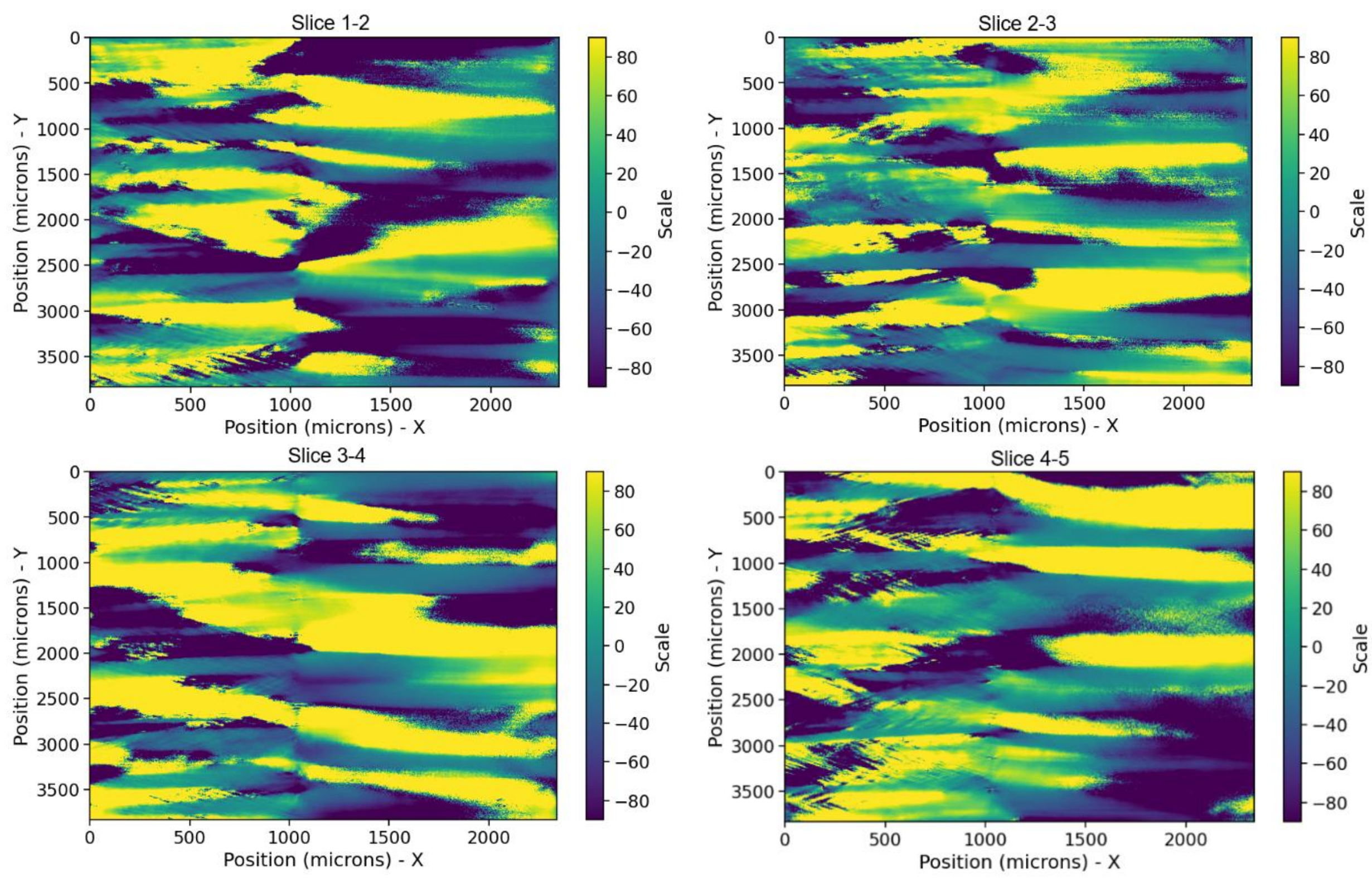

*Figure S4: Azimuth maps of the Slices used for calculating correlation coefficients between layers parallel to the growth direction*

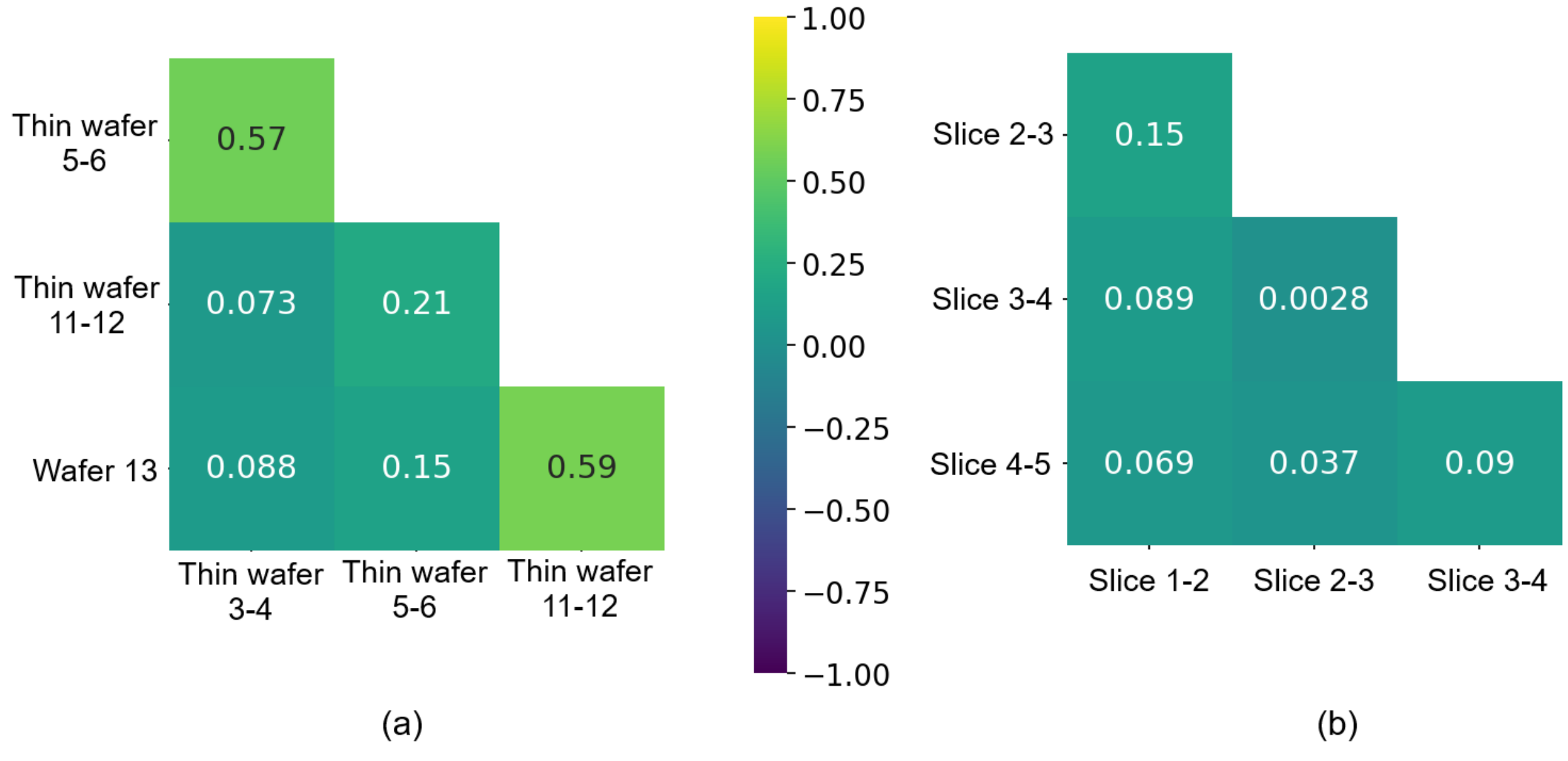

*Figure S5: Correlation coefficients of retardance of some available (a) Thin Wafers and (b) Slices*

*Table S1: Cross samples and their corresponding thicknesses and mean retardances*

| Sample | Thickness (µm) | Mean retardance (Degrees) |
|---|---|---|
| Cross1 | 2040 | 11.14 |
| Cross2 | 1760 | 10.2 |
| Cross3 | 1500 | 8.79 |
| Cross4 | 1240 | 8.64 |
| Cross5 | 985 | 6.72 |
| Cross6 | 780 | 6.61 |
| Cross7 | 588 | 6.03 |
| Cross8 | 429 | 4.78 |
| Cross9 | 286 | 3.29 |
| Cross10 | 120 | 2.33 |

*Table S2: Thickness and mean retardance of Slices cut out of Cross samples used for analyzing correlation*

| Sample | Thickness (µm) | Mean retardance (Degrees) |
|---|---|---|
| Slice 1-2 | 195 | 3.58 |
| Slice 2-3 | 205 | 3.22 |
| Slice 3-4 | 180 | 4.12 |
| Slice 4-5 | 145 | 3.03 |

*Table S3: Wafer samples and their corresponding thickness and mean retardance*

| Sample | Thickness (µm) | Mean retardance (Degrees) |
|---|---|---|
| Wafer 1 | 2078 | 65.4 |
| Wafer 2 | 1930 | 62.5 |
| Wafer 3 | 1755 | 57.3 |
| Wafer 4 | 1570 | 56.5 |
| Wafer 5 | 1420 | 51.79 |
| Wafer 6 | 1234 | 44.59 |
| Wafer 7 | 1070 | 38.16 |
| Wafer 8 | 973 | 35.49 |
| Wafer 9 | 810 | 29.99 |
| Wafer 10 | 629 | 23.39 |
| Wafer 11 | 456 | 17.49 |
| Wafer 12 | 276 | 10.64 |
| Wafer 13 | 74 | 2.89 |

*Table S4: Thickness and mean retardance of thin Wafers cut out of Cross samples used for analyzing correlation*

| Sample | Thickness (µm) | Mean retardance (Degrees) | Base retardance (Degrees/ µm) |
|---|---|---|---|
| Thin wafer 3-4 | 80 | 3.52 | 0.044 |
| Thin wafer 5-6 | 74 | 3.30 | 0.045 |
| Thin wafer 11-12 | 65 | 2.73 | 0.042 |
| Wafer 13 | 74 | 2.92 | 0.039 |

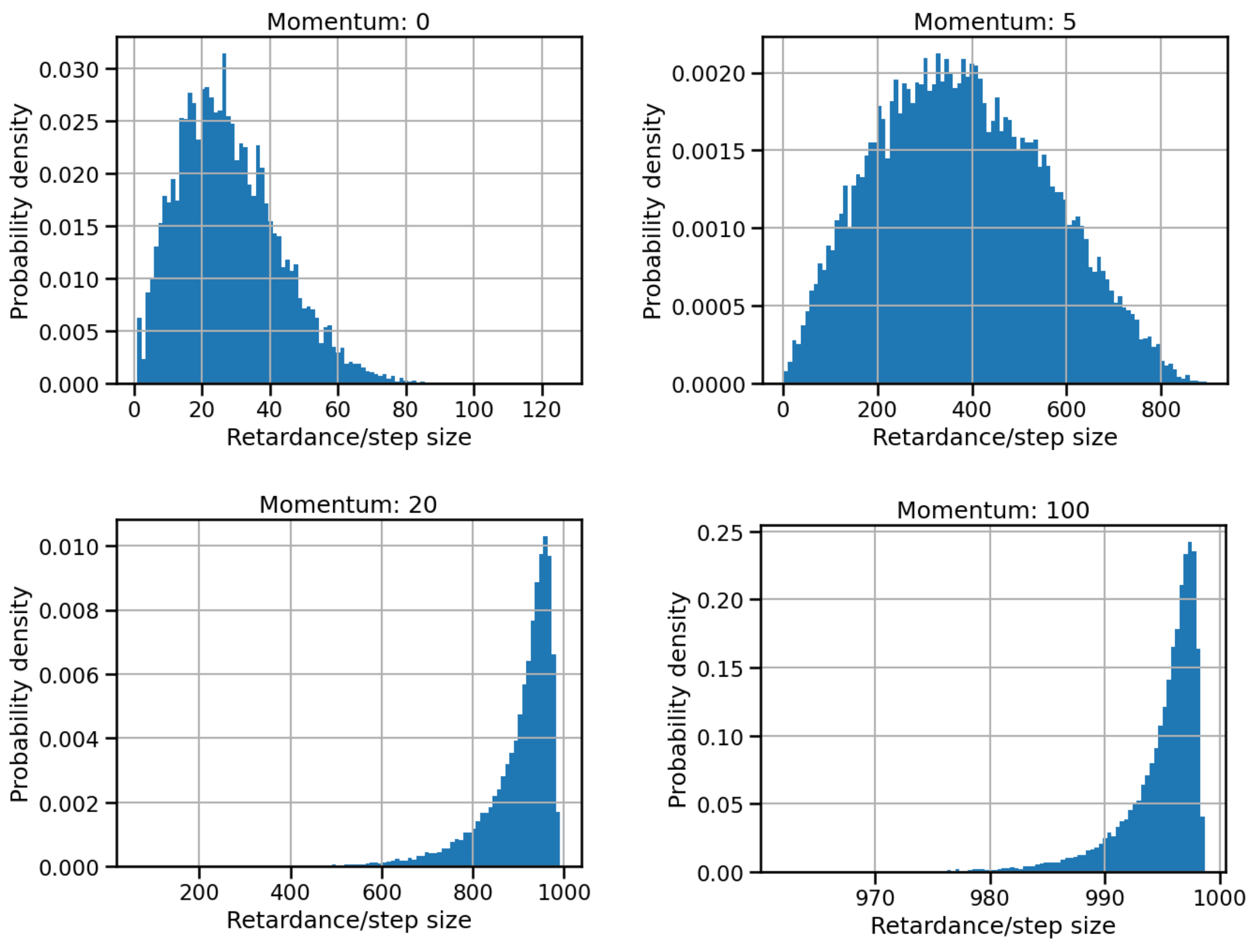

*Figure S6: Probability density function of final retardance of 2000 momentum drift random walk after 1000 steps at momentum factors of 0, 5, 20, and 100*

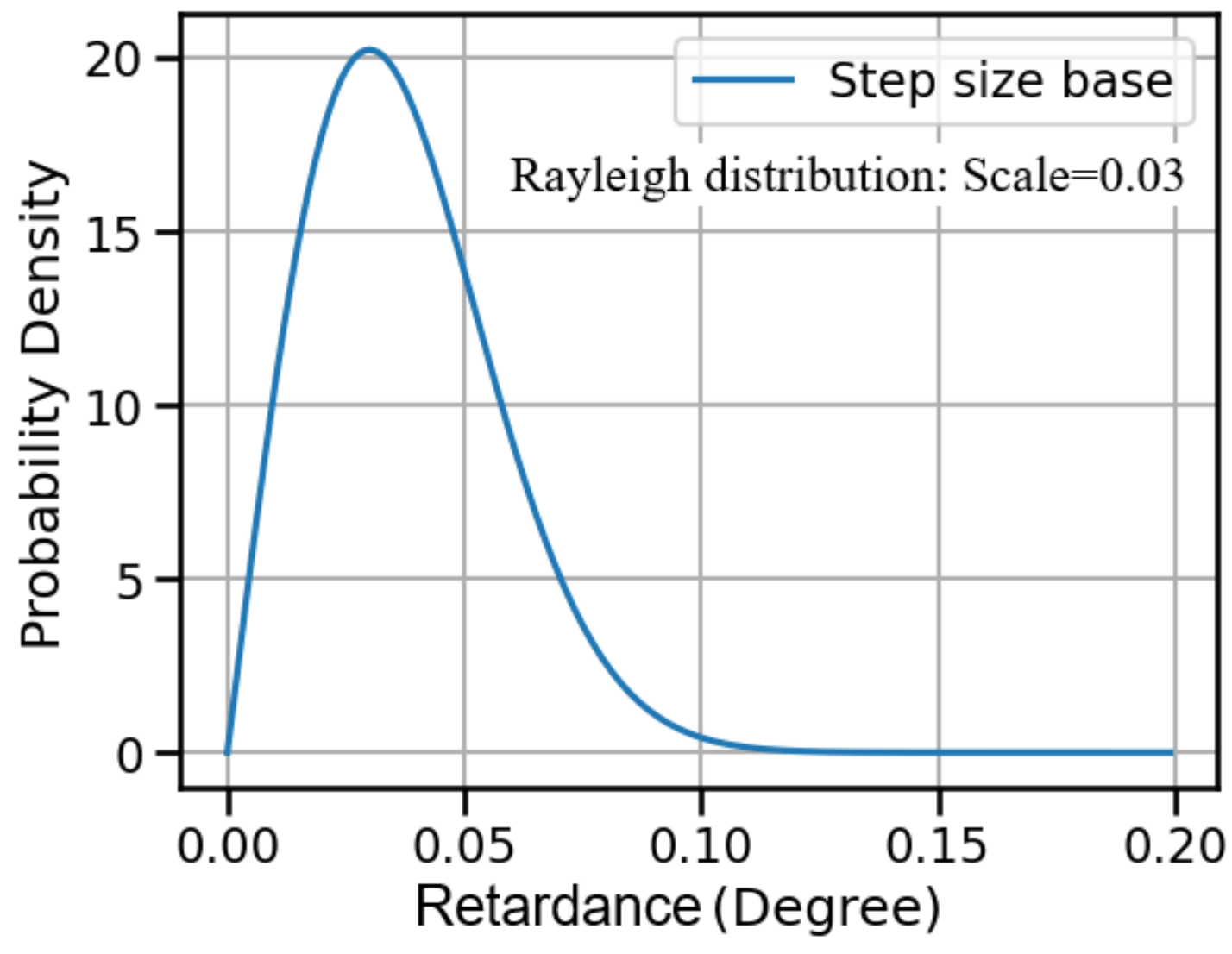

*Figure S7: Probability density function of base retardance of 1 μm step layer*

Retardance probability distribution function of a unit layer having the thickness of one step. It is worth noticing that space step size is kept constant at 1 $\mu m$ and retardance step size varies and it follows Rayleigh distribution with the scale factor of 0.03 chosen in this work:

$$PDF_{base}(x,\sigma) = \frac{x}{\sigma^2} e^{-x^2/(2\sigma^2)}, x \geq 0$$

with $x$ is retardance step size and $\sigma$ is scale factor. $\sigma = 0.03$ in this work because $Mean = \sigma\sqrt{\frac{\pi}{2}} \approx 0.04$ , which is close to experimental estimates.

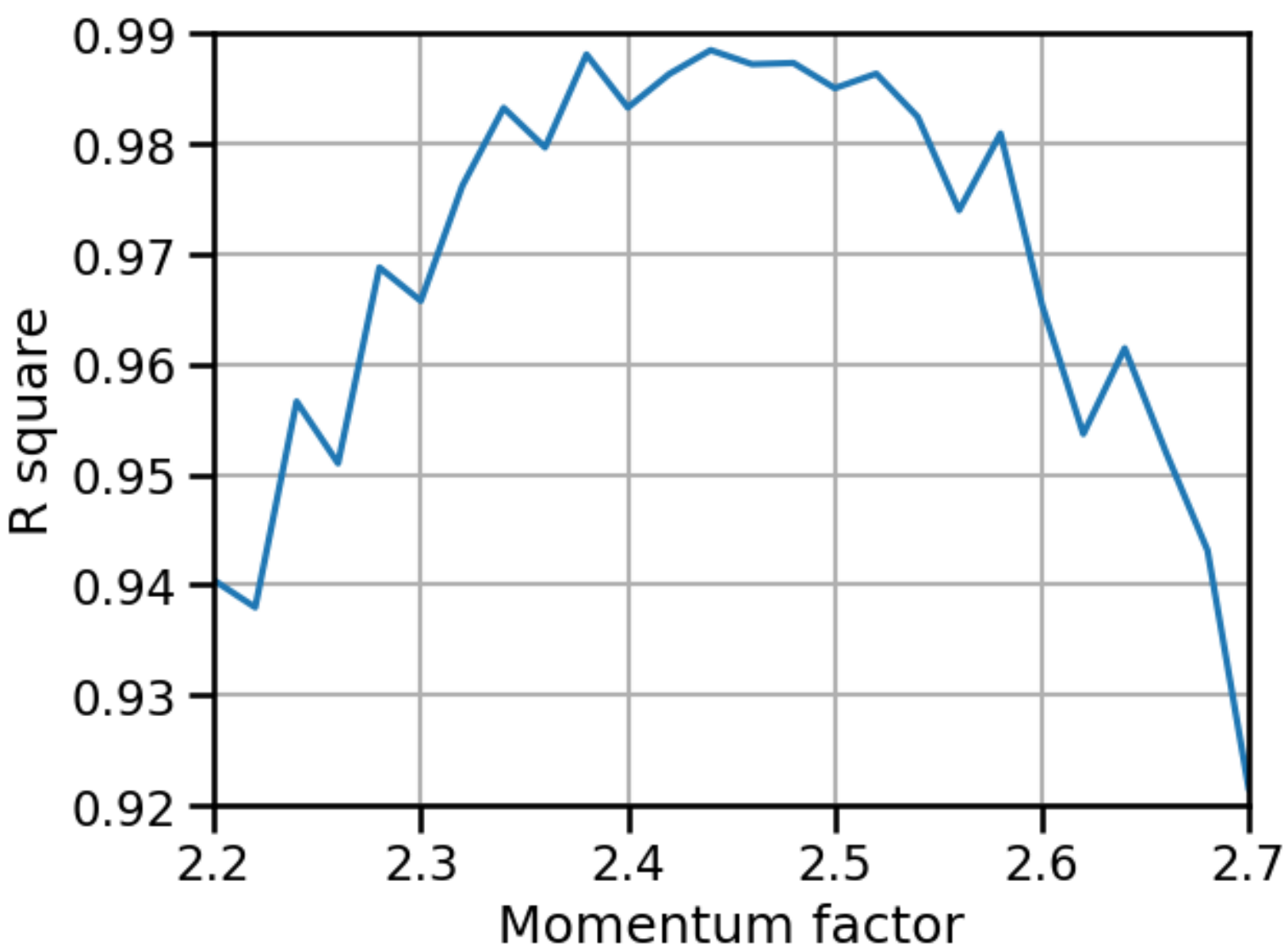

*Figure S8: Optimizing R square for momentum factor of the Cross samples*

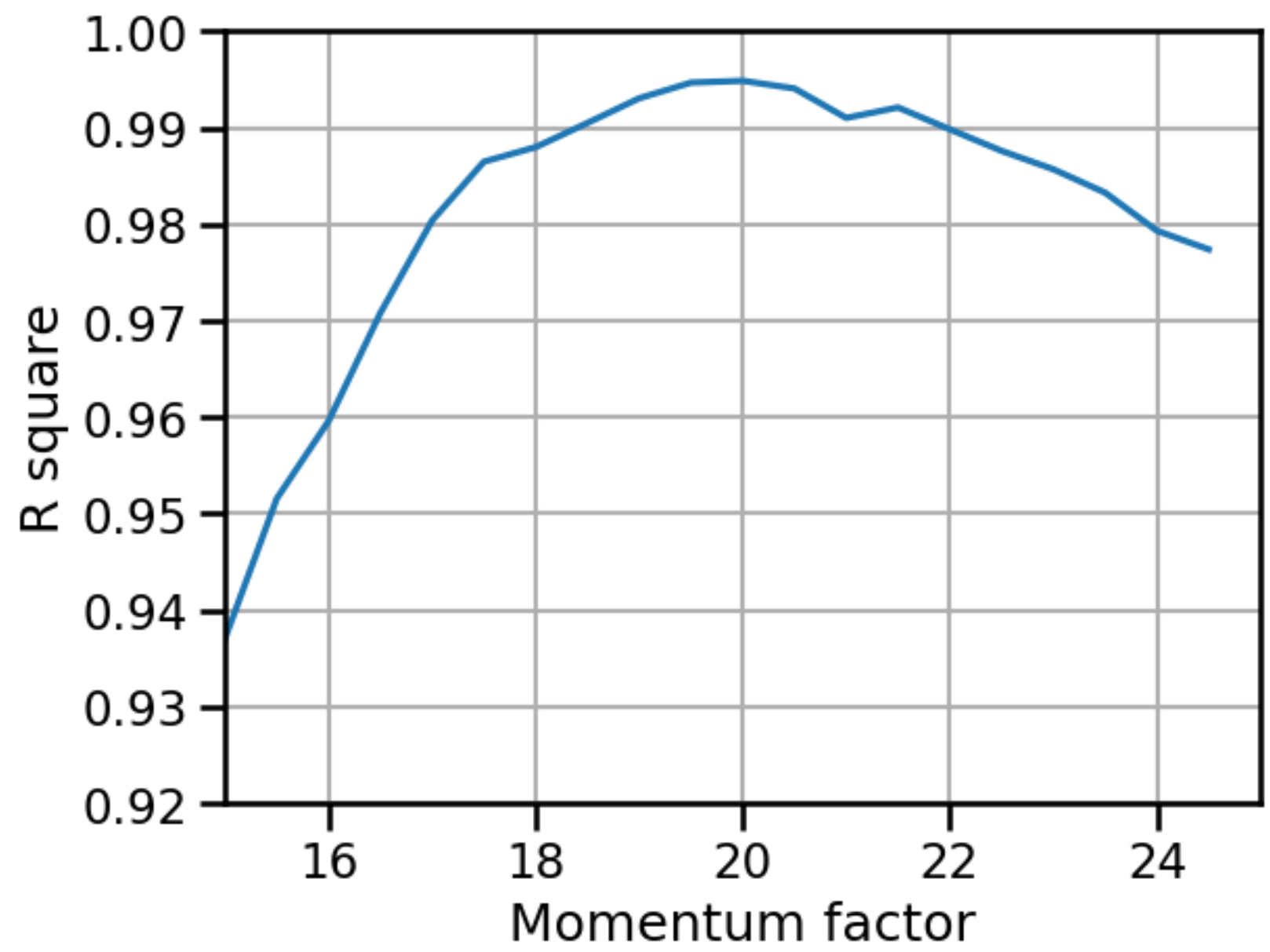

*Figure S9: Optimize R square for momentum factor of Wafer samples*

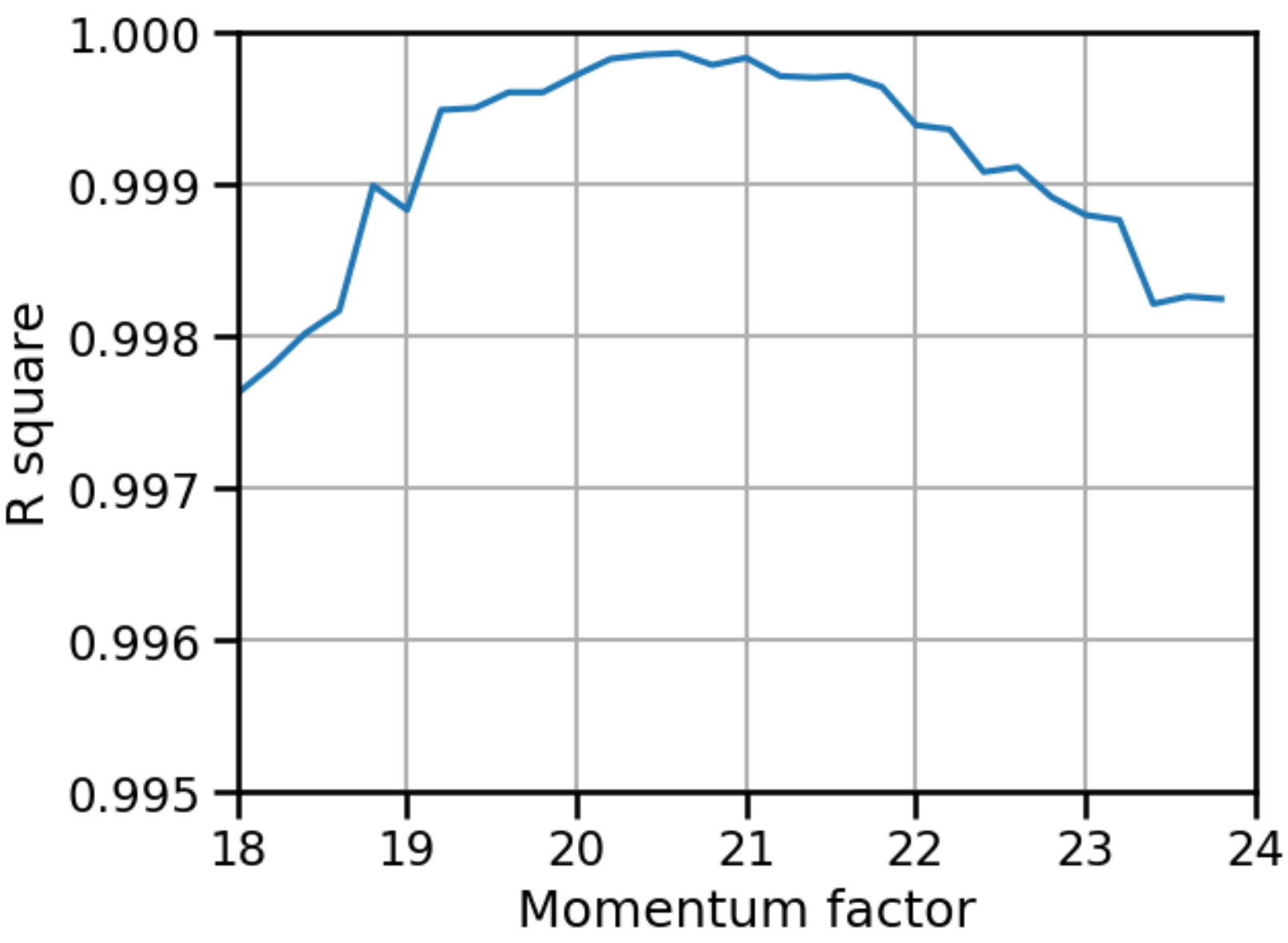

*Figure S10: Optimizing R square for momentum factor of the seed layer*

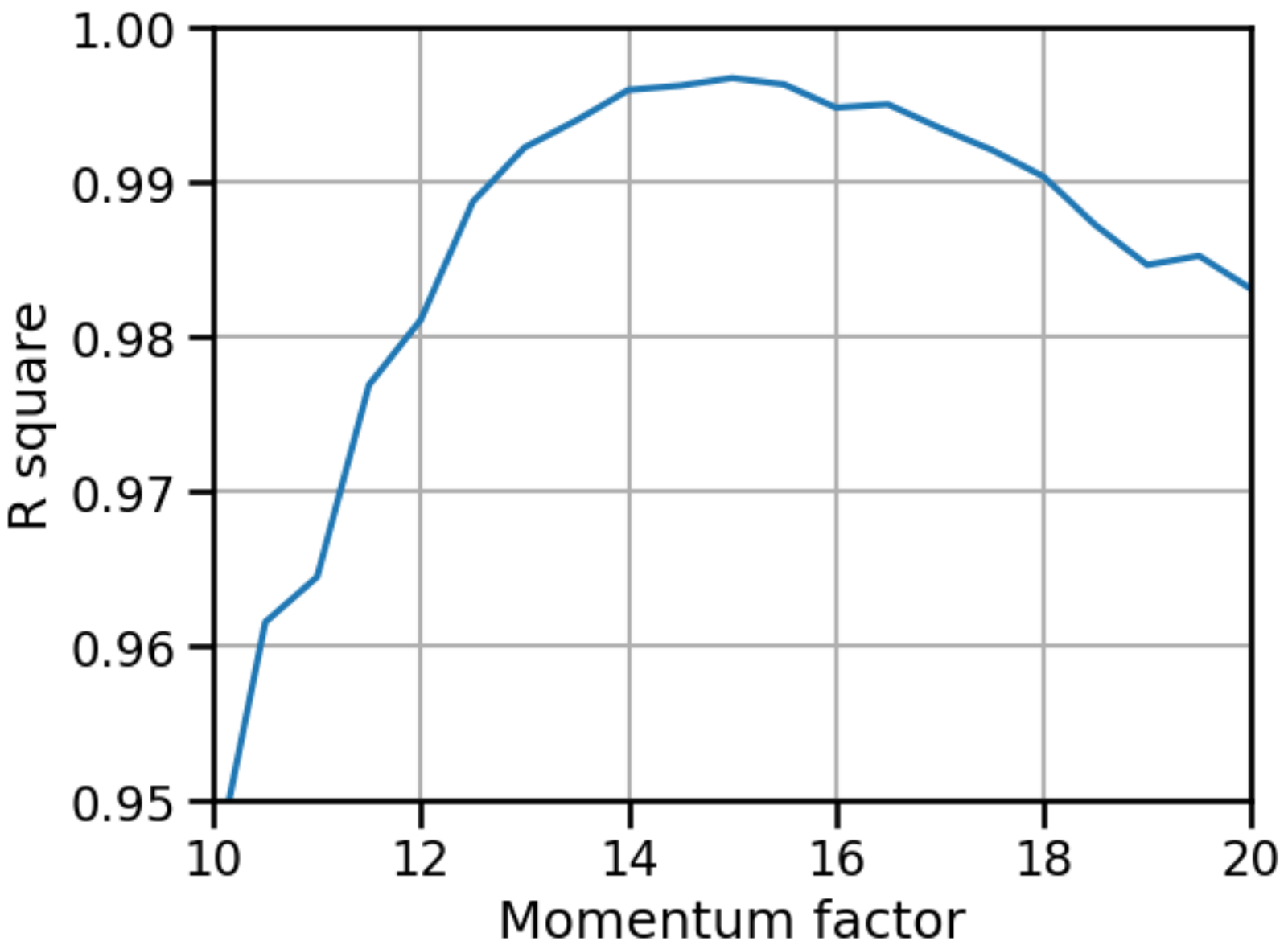

*Figure S11: Optimizing R square for momentum factor of the grown layer*